\documentclass[12pt,preprint]{aastex}

\usepackage{wrapfig}
\usepackage{amsmath}
\usepackage{amsfonts}
\usepackage{amssymb}

\def\lapprox{\mathrel{\hbox{\rlap{\hbox{\lower4pt\hbox{$\sim$}}}\hbox{$<$}}}}
\def\gapprox{\mathrel{\hbox{\rlap{\hbox{\lower4pt\hbox{$\sim$}}}\hbox{$>$}}}}

\newcommand{\be}{\begin{equation}}
\newcommand{\ee}{\end{equation}}

\newcommand {\nind} {\noindent}
\newcommand {\mb} {\mathbf}
\newcommand {\mbb} {\mathbb}
\newcommand {\bea} {\begin{eqnarray}}
\newcommand {\eea} {\end{eqnarray}}
\newcommand {\ve} {\varepsilon}

\begin{document}

\shorttitle{Chromospheric reconnection}
\shortauthors{Leake et al.}

\title{Multi-fluid simulations of chromospheric magnetic reconnection in a weakly ionized reacting plasma}

\author{James E. Leake}
\affil{College of Science, George Mason University, 4400 University Drive, Fairfax, Virginia 22030. \\
jleake@gmu.edu}

\author{Vyacheslav S. Lukin, Mark G. Linton}
\affil{U.S. Naval Research Lab 4555 Overlook Ave., SW Washington, DC 20375.}

\and
\author{Eric T. Meier}
\affil{Lawrence Livermore National Laboratory 7000 East Avenue,  Livermore, CA 94550. }

\begin{abstract}
We present results from the first self-consistent multi-fluid simulations of chromospheric magnetic reconnection in a weakly ionized reacting plasma. We simulate two dimensional magnetic reconnection in a Harris current sheet with a numerical model which includes ion-neutral scattering collisions, ionization, recombination, optically thin radiative loss,  collisional heating, and thermal conduction. 
In the resulting tearing mode reconnection the neutral and  ion fluids
become decoupled upstream from the reconnection site, creating an
excess of ions  in the reconnection region and therefore an ionization
imbalance. Ion recombination in the reconnection region, combined with Alfv\'{e}nic outflows, quickly removes ions from the reconnection site, leading to a fast reconnection rate independent of Lundquist number. In addition to allowing fast reconnection, we find that these non-equilibria partial ionization effects lead to the onset of the nonlinear secondary tearing instability at lower values of the Lundquist number than has been found in fully ionized plasmas.
  These simulations provide evidence that magnetic reconnection in the chromosphere could be responsible for jet-like transient phenomena  such as spicules and chromospheric jets.
\end{abstract}

\section{Introduction}

Magnetic reconnection is the general process by which free magnetic energy stored in a plasma can be converted into kinetic and thermal energy by breaking the frozen-in constraint which exists for a perfectly conducting plasma. Magnetic reconnection occurs in a variety of astrophysical plasmas, including the interstellar medium (ISM), galactic disks, and the solar atmosphere \citep{2009ARA&A..47..291Z}.  It has been suggested as a cause of many transient phenomena, such as solar flares and X-ray jets \citep{2011ApJ...731L..18M}, magnetospheric substorms \citep{2007JGRA..11206215B,2007JGRA..11201209M} and solar $\gamma$-ray bursts \citep{2005JGRA..11011103E,2005ApJ...628L..77T}. Reconnection also occurs in laboratory plasmas \citep{2010PhPl...17j2106G}, and plays a key role in the self-organization of fusion plasmas \citep{2006PhRvL..96s5003P,2007PhRvL..98g5001G}. 

Magnetic reconnection has long been considered as a mechanism for creating transient phenomena in the corona, such as solar flares, coronal mass ejections, and more recently X-ray jets (see review by \citet{2011ApJ...731L..18M}). In addition, recent observations of the solar chromosphere, the cooler, weakly ionized plasma below the solar corona, have suggested that reconnection is occurring there also. In particular, the quiet chromosphere exhibits localized, transient outflows on a number of different length-scales. The largest class of such outflows are called  ``chromospheric jets''  \citep{2007Sci...318.1591S}.
These are surges of plasma, observed in H$\alpha$ and Ca-II, with typical lifetimes of 200-1000 s, lengths of 5 Mm, and velocities at their base of 10 km/s. Analysis of observations of chromospheric jets with Hinode have shown ``blobs" of plasma within the jet outflow \citep{2007Sci...318.1591S}. 
A smaller class of these outflows are called ``spicules" \citep{2000SoPh..196...79S}, and  are mainly observed on the solar limb, though possible disk counterparts have recently been identified in blue-shifts of Ca I and H$\alpha$ emission \citep{2012ApJ...752..108S}. 
Spicules have lifetimes of 10-600 s, lengths of up to 1 Mm and velocities of 20-150 km/s.

A unified model of chromospheric outflow generation has recently been suggested by
 \citet{2007Sci...318.1591S} and \citet{2011ApJ...731L..18M}. The basic structure of the theoretical model consists of the so-called ``anemone" structure: a bipole field emerging into and reconnecting with a unipolar field. By changing the size of the emerging bipole, this simple model has been presented as a way of explaining spicules and chromospheric jets, as well as solar X-ray jets. Other evidence that magnetic reconnection is a possible driver of chromospheric transient phenomena includes Alfv\'{e}nic flows in spicules and ``blobs" of plasma in chromospheric jets, both of which are by-products of magnetic reconnection. 


A naive estimate of the reconnection rate $M$ (the rate at which magnetic field reconnects and ejects plasma) for these chromospheric phenomena can be made by dividing  the inflow rate of plasma into the reconnection site by the outflow rate. The simple ``anemone'' structure described above has a flux inflow reservoir which is of the same size as the outflow region, and when this reservoir of flux is depleted, the reconnection stops. Therefore a typical size $L$ is used for both inflow and outflow regions, and  given an outflow speed $v_{out}$, and a lifetime $t_{life}$,  the reconnection rate can be estimated as
$M\approx\frac{L }{v_{out}t_{life}}$. Taking the range of lifetimes, lengths and velocities for chromospheric jets gives a minimum of $M\approx 0.5$. Doing the same for spicules gives a minimum of $M\approx 0.01$.

Traditionally, magnetic reconnection has been considered within the single-fluid, fully ionized, magnetohydrodynamic (MHD) framework, which is applicable to collisional plasmas. For a highly conducting plasma, the diffusion due to electron-ion collisions is negligible and the magnetic fieldlines are frozen into the plasma. Reconnection occurs when the frozen-in constraint is broken on timescales much shorter than the classical diffusion time, by allowing fieldlines to reconnect through a narrow diffusion region. \citet{1957JGR....62..509P} and \citet{Sweet58} were the first to formulate magnetic reconnection as a local process by considering a current layer of width $\delta$ much smaller than its length $L$, and with non-zero resistivity ($\eta$) due to electron-ion collisions. The model assumes steady state, i.e.,  the 
rate of plasma flowing into the diffusion region is equal to the rate of plasma flowing out, and takes the length $L$ to be the characteristic system length scale.  Using Ohm's law for a fully ionized, single fluid plasma
\be
\mathbf{E}+\mathbf{v}\times\mathbf{B}  = \eta \mathbf{j},
\ee
and using the plasma momentum equation, a simple scaling law for the reconnection rate can be derived:
\be
M \equiv \frac{v_{in}}{v_{A}} \approx \sqrt{\frac{\eta}{\mu_{0}v_{A}L}} = \frac{1}{\sqrt{S}}.
\ee
Here $S=\mu_{0}v_{A}L/\eta$ is the Lundquist number, and $v_{A}=B/\sqrt{\rho\mu_{0}}$ is a typical Alfv\'{e}n velocity, with $B$ evaluated upstream from the current sheet and $\rho$ evaluated in the current sheet. From this analysis, it can also be shown that the current sheet aspect ratio $\sigma \equiv \frac{\delta}{L}$ scales as
\be
\sigma \propto \frac{1}{\sqrt{S}}.
\ee

Using the one dimensional (1D) semi-implicit model for the quiet Sun of
\citet{1981ApJS...45..635V} gives a mass density of $7.4\times10^{-8}
~ \textrm{kg}/\textrm{m}^3$ for the chromosphere at 1 Mm above
the solar surface. Using a magnetic field strength of 50 G gives an
Alfv\'{e}n speed of approximately 15 km/s. The temperature at 1 Mm in
this 1D model is  $\approx$ 6000 K and the electron density is $\approx
10^{17} ~ \textrm{m}^{-3}$, which gives the electron-ion collision
frequency to be  $10^7 ~ \textrm{s}^{-1}$, and the Spitzer resistivity
to be 0.004 $\Omega$m. Assuming a typical length scale of 0.1 Mm gives
a Lundquist number of $S\approx10^6$ and a Sweet-Parker  reconnection
rate of $M\approx10^{-3}$. This is slower than the $M \ge 0.01$ required if reconnection is to explain the observed lifetimes of both chromospheric jets and spicules. 
The problem is greater in the corona, where $S\approx 10^{9}$, and the Sweet-Parker model predicts reconnection rates much too slow relative to those implied from observations. This is the general problem of the Sweet-Parker model: the predicted reconnection rate is significantly slower than almost all atmospheric transient phenomena.
 
\citet{Petschek64} modified the Sweet-Parker reconnection formulation by allowing plasma to be redirected by standing shock waves which are setup at the ends of the diffusion region. This allowed for a shorter diffusion region, with length $L'$, and the reconnection rate increased by $\sqrt{L/L'}$. The maximum reconnection rate was found to be 
$M = \frac{\pi}{8 \ln(S)}$.
For a value of $S=10^6$, this gives 0.065: substantially faster than the Sweet-Parker prediction of 0.001, and within the range observed for chromospheric spicules, though still too small for chromospheric jets.

While the Petschek model predicts faster reconnection rates than the Sweet-Parker prediction,  numerical simulations can only reproduce the Petschek reconnection regime if the resistivity is localized near the X-point - so-called \textit{anomalous resistivity} due to turbulence in the current layer or ion-cyclotron wave effects \citep{2003ApJ...587..450U,2004PhPl...11.2199S} - or if ion-electron drift terms are retained in the Ohm's law \citep{Birn01}.


More recently, \citet{Loureiro07} and \citet{Huang10} considered the secondary tearing  instability of the Sweet-Parker current sheet in a fully ionized plasma as an alternative means for obtaining fast reconnection.  When the current sheet aspect ratio ($\sigma \equiv \delta/L$) reached a critical value of $\sigma_{c} = 1/200$, the sheet became unstable to the secondary tearing instability. Here, thinner current sheets were created between the primary plasmoids, which were themselves then prone to breaking into secondary plasmoids. With Sweet-Parker scaling in the laminar regime, they found that  secondary onset occurred for a critical Lundquist number $S_{c}$ of $S_{c} = 1/\sigma_{c}^{2}= 4\times10^{4}$. Above this value, the reconnection rate was independent of $S$: $M=1/\sqrt{S_{c}}$. For a chromospheric Lundquist number of $S=10^{6}$, the reconnection rate as a result of the plasmoid instability gives $M=1/\sqrt{S_{c}}=5\times10^{-3}$, 5 times the Sweet-Parker prediction.
Thus, the secondary instability allows reconnection independent of the mechanism which breaks the frozen-in constraint, but still gives reconnection rates slower than is needed to explain the parameters associated with chromospheric reconnection events.  While secondary tearing may be the cause of the plasma ``blobs'' seen in chromospheric jets, until now no simulations have been performed to see if chromospheric magnetic reconnection could yield these plasmoids. 

This paper studies magnetic reconnection in the partially ionized chromosphere, focusing on both fast laminar reconnection and plasmoid formation.  The plasma-$\beta$, the ratio of plasma pressure to magnetic pressure,  can be as high as $10^{2}$ and as low as $10^{-4}$ in this region of the solar atmosphere  \citep{2001SoPh..203...71G}. The average mass density falls over 4 orders of magnitude in a few Mm \citep{1981ApJS...45..635V} from the photosphere to the base of the corona. The 
ionization fraction, defined as the percentage of the plasma which is
ionized, ranges from 0.1\% to 50\%, as the neutral density falls off
faster than the ionized component density. In the bulk of the
chromosphere the average collision time between neutral atoms and
ions is of the order of ms \citep{1981ApJS...45..635V}. This is much
less than a typical chromospheric time scale of 7~s, based on a typical Alfv\'{e}n velocity of 15~km/s and a typical length scale of 100~km. Hence the ions and neutrals are often treated as a single fluid. However, when magnetic diffusion length scales become as small as the neutral-ion collision mean free path,  e.g., at magnetic reconnection sites, the decoupling of neutrals and ions cannot necessarily be neglected, and chromospheric magnetic reconnection should be studied in a multi-fluid framework.

Partial ionization affects magnetic reconnection in several ways. The
most obvious is the effect on the resistivity. Ion-neutral collisions
introduce a Pedersen resistivity (or equivalently, an ambipolar
diffusion) in the single-fluid MHD formulation, which acts
perpendicular to the magnetic field \citep{1965RvPP....1..205B}.
\citet{1994ApJ...427L..91B,1995ApJ...448..734B} showed that this
ambipolar dissipation can create thin current structures in weakly
ionized plasmas. The ionization level can also affect the reconnection rate through the Alfv\'{e}n speed: for coupled systems, the outflow Alfv\'{e}n speed depends on the total plasma density, but for decoupled systems it depends only on the ionized component density.

Previously, \citet{1999ApJ...511..193V}, \citet{2003ApJ...583..229H}, and 
  \citet{2004ApJ...603..180L} considered magnetic reconnection in the weakly
ionized ISM, using a 1D analytic approach (note that the 1D paradigm
assumes that outflow is negligible).  In the reconnection inflow, if
the ions that are pulled in by the reconnecting magnetic field are
decoupled from the neutrals, an excess of ions can build up in the
reconnection region. Recombination can then act as a sink for the
ions, also decoupling them from the field. Given a sufficiently large
recombination sink, the reconnection rate can become independent of
the resistivity. This prediction has not been studied in higher
dimensional magnetic reconnection configurations. Below, we examine
this and other aspects of magnetic reconnection in a weakly ionized
plasma within a two-dimensional (2D) numerical model.

\citet{2008A&A...486..569S} and \citet{2009ApJ...691L..45S} performed two-fluid (ion+neutral) simulations of the coalescence of magnetic structures in partially ionized plasmas. They found that the reconnection rate for a fixed resistivity decreased as the plasma became less ionized, suggesting that jets associated with fast reconnection must occur in the upper chromosphere, where the ionization level is higher.
However, in their investigations, the collision frequency and ionization/recombination rates did not depend self-consistently on the local plasma parameters (temperature and density), and  therefore did not vary with space and time during the magnetic reconnection, a phenomena which is vital to test the predictions of  \citet{1999ApJ...511..193V}, \citet{2003ApJ...583..229H}, and 
  \citet{2004ApJ...603..180L}. In addition, their choice of resistivity was approximately 10 $\Omega$m, which is 4 orders of magnitude larger than a typical value in the chromosphere, and their Lundquist number was subsequently a relatively low value of approximately 10.

In this paper, the first self consistent 2D simulations of chromospheric magnetic reconnection in a weakly ionized reacting plasma are performed. The numerical model used to simulate this reconnection is presented in \S2. The results are presented in \S 3, with particular focus on the scaling of the reconnection rate with resistivity, and the onset of the plasmoid instability. In \S 4 these results are used to evaluate the likelihood that magnetic reconnection can explain transient phenomena observed in the chromosphere such as spicules and jets.

\section{Numerical Method}

\subsection{Multi-Fluid Partially Ionized Plasma Model}
\label{sec:HiFi_PN}

A partially ionized reacting multi-fluid hydrogen plasma model is used to simulate reconnection in the solar chromosphere. For a detailed description and derivation of the model the reader is directed to \citet{Meier11} and \citet{Meier12b}. The model is implemented in the implicit, adaptive high order finite (spectral) element code framework, HiFi \citep{2008PhDT.........1L}. 

The full model consists of three fluids, ion (\textit{i}),
electron (\textit{e}), and neutral (\textit{n}).  The fluids can undergo  
recombination, ionization and charge exchange interactions, with
 $\Gamma_\alpha^r$  denoting the reaction rate for interaction $r$ affecting fluid
$\alpha$. 

The recombination reaction rate for ions (the rate of change of ion number density due to recombination), $\Gamma_{i}^{rec}$, is defined as 
\be
\Gamma_{i}^{rec} \equiv -n_{i}\nu^{rec},
\ee
where the recombination frequency 
\be
\nu^{rec} = n_{e}\frac{1}{\sqrt{T_{e}^{*}}} 2.6\times10^{-19} \textrm{m}^{3}\textrm{s}^{-1}
\label{eqn:recomb}
\ee
is obtained from \citet{2003poai.book.....S} and $T_{e}^{*}$ is the
electron temperature $T_e$ specified in eV.

The ionization reaction rate for neutrals, $\Gamma_{n}^{ion}$, is defined as
\be
\Gamma_{n}^{ion} \equiv  -n_{n}\nu^{ion},
\label{eqn:ioniz}
\ee
where the ionization frequency
\be
\nu^{ion} = n_{e}A \frac{1}{X+\phi_{ion}/T_{e}^{*}}\left(\frac{\phi_{ion}}{T_{e}^{*}}\right)^{K}e^{-\phi_{ion}/T_{e}^{*}} \textrm{m}^{3}\textrm{s}^{-1}
\ee
is given by the practical fit from \citet{1997ADNDT..65....1V}, using the values $A=2.91\times10^{-14}$, $K=0.39$, $X=0.232$, and the Hydrogen ionization potential $\phi_{ion}=13.6 \textrm{eV}$. Note that $\Gamma_{i}^{ion} = -\Gamma_{n}^{ion}$, and $\Gamma_{n}^{rec} = -\Gamma_{i}^{rec}$.

The charge exchange reaction rate, $\Gamma^{cx}$ is defined as \be
\Gamma^{cx} \equiv \sigma_{cx}(V_{cx})n_{i}n_{n}V_{cx} \ee where \be
V_{cx}\equiv
\sqrt{\frac{4}{\pi}v_{Ti}^{2}+\frac{4}{\pi}v_{Tn}^{2}+v_{in}^{2}} \ee
is the representative speed of the interaction and $v_{in}^{2} \equiv
{|\mb{v}_{i}-\mb{v}_{n}|}^2$ with $\mb{v}_{\alpha}$ denoting the velocity of
species $\alpha$. The thermal speed of species $\alpha$ is given by
$v_{T\alpha} = \sqrt{\frac{2k_{B}T_{\alpha}}{m_{\alpha}}}$, where
$T_{\alpha}$ is the temperature, $m_{\alpha}$ is the corresponding
particle's mass, and $k_{B}$ is Boltzmann's constant.  Functional
forms for the charge exchange cross-section $\sigma_{cx}(V_{cx})$ can be found in \citet{Meier11}.

The multi-fluid model can be expressed by taking the neutral continuity
equation, momentum equation, and energy (or pressure) equation to obtain a set of equations for
the neutral fluid, and combining the electron and ion versions of these equations to 
obtain a set of equations for the ``ionized'' fluid. 
These equations are supplemented with Faraday's Law, and the required transport
closure equations. Assuming charge neutrality in a hydrogen plasma, the 
 electron and ion number densities are set
equal to each other ($n_{i}=n_{e}$). Also the 
ionized and neutral atom masses are set equal to the proton mass
($m_{i}=m_{n}=m_{p}$). The resulting system of partial differential equations (PDEs) is given below.

\nind\textit{Continuity: } \\
Due to charge neutrality, only the ion and neutral continuity equations are required.
\bea
 \frac{\partial n_{i}}{\partial t} + \nabla \cdot (n_{i} \mb{v}_i) & =
& \Gamma_i^{ion} + \Gamma_i^{rec}, \\
 \frac{\partial n_n}{\partial t} + \nabla \cdot (n_n \mb{v}_n) & 
= & \Gamma_n^{rec} + \Gamma_n^{ion}. \eea

\nind\textit{Momentum: } \\
The electron and ion momentum equations are summed and terms of order $(m_e/m_p)^{1/2}$ and higher are neglected to give:
\bea
\label{eq:mom_i}
\frac{\partial}{\partial t} (m_i n_{i} \mb{v}_i) + 
\nabla \cdot (m_i n_{i} \mb{v}_i \mb{v}_i + \mathbb{P}_i+ \mathbb{P}_e) & = &
 \mb{j} \times \mb{B} + \mb{R}_i^{in} 
+ \Gamma_i^{ion} m_i \mb{v}_n -
\Gamma_n^{rec} m_i \mb{v}_i \nonumber \\
 & +&  \Gamma^{cx} m_i (\mb{v}_n - \mb{v}_i) +
\mb{R}_{in}^{cx} - \mb{R}_{ni}^{cx}. \eea

\nind The neutral momentum equation is
\bea
\label{eq:mom_n}
\frac{\partial}{\partial t} (m_i n_n \mb{v}_n) +  \nabla \cdot (m_i
n_n \mb{v}_n \mb{v}_n + \mathbb{P}_n) & = & - \mb{R}_i^{in} 
+  \Gamma_n^{rec} m_i \mb{v}_i
-\Gamma_i^{ion} m_i \mb{v}_n \nonumber \\
& + &  \Gamma^{cx} m_i (\mb{v}_i -
\mb{v}_n) - \mb{R}_{in}^{cx} + \mb{R}_{ni}^{cx}. \eea 

\nind Here the current density is
 \be
 \mathbf{j} = en_{i}(\mathbf{v}_{i}-\mathbf{v}_{e})=\frac{\nabla\times\mb{B}}{\mu_{0}}.
 \ee
The momentum transfer  $\mb{R}_\alpha^{\alpha\beta}$ is the transfer
of momentum to species $\alpha$ due to identity-preserving collisions
with species $\beta$: 
\be
\mb{R}_\alpha^{\alpha\beta} = m_{\alpha\beta}n_{\alpha}\nu_{\alpha\beta}(\mb{v}_{\beta}-\mb{v}_{\alpha}),
\ee
where $m_{\alpha\beta}  = \frac{m_{\alpha}m_{\beta}}{m_{\alpha}+m_{\beta}}$. The collision frequency $\nu_{\alpha\beta}$ is given by
\be
\nu_{\alpha\beta} = n_{\beta}\Sigma_{\alpha\beta}\sqrt{\frac{8k_{B}T_{\alpha\beta}}{\pi m_{\alpha\beta}}},
\ee
with $T_{\alpha\beta} = \frac{T_{\alpha}+T_{\beta}}{2}$. The cross-section $\Sigma_{in} = \Sigma_{ni}$ is $1.41\times 10^{-19} ~ \textrm{m}^2$, and the cross-section
$\Sigma_{en} = \Sigma_{ne}$ is $1\times 10^{-19} ~ \textrm{m}^2$, assuming solid sphere elastic collisions \citep{1983ApJ...264..485D}.
The cross section for ion-electron collisions, $\Sigma_{ei}= \Sigma_{ie}$, is assumed to be $\pi r_{d}^{2}$ where $r_{d}$ is the distance of closest approach ($e^{2}/(4\pi\epsilon_{0} k_{B}T_{ei})$ with $\epsilon_{0}$ the permittivity of free space and $e$ the elementary charge).

The momentum transfer from species $\beta$ to species $\alpha$ due to charge exchange is $\mb{R}_{\alpha\beta}^{cx}$.
As found by \citet{1995JGR...10021595P} and detailed in \citet{Meier11} and \citet{Meier12b}, appropriate approximations for these terms are
\be
\mb{R}_{in}^{cx} \approx -m_{i}\sigma_{cx}(V_{cx})n_{i} n_{n}\mb{v}_{in}v_{Tn}^{2}\left[4\left(\frac{4}{\pi}v_{Ti}^{2}+v_{in}^2 \right)+\frac{9}{4\pi}v_{Tn}^{2}\right]^{-1/2},
\ee
and
\be
\mb{R}_{ni}^{cx} \approx  m_{i}\sigma_{cx}(V_{cx})n_{i} n_{n}\mb{v}_{in}v_{Ti}^{2}\left[4\left(\frac{4}{\pi}v_{Tn}^{2}+v_{in}^2 \right)+\frac{9}{4\pi}v_{Ti}^{2}\right]^{-1/2}.
\ee

The pressure tensor is 
 $\mbb{P}_\alpha = P_\alpha\mbb{I} + \pi_\alpha$ where $P_\alpha$ is
the scalar pressure and $\pi_\alpha$ is the viscous stress tensor, given by 
$\pi_{\alpha} = -\xi_{\alpha}[\nabla\mb{v}_{\alpha} + (\nabla\mb{v}_{\alpha})^{\top}]$ where $\xi_{\alpha}$ is the isotropic dynamic viscosity coefficient for the fluid $\alpha$.

\nind\textit{Internal Energy:} \\
Again, combining the electron and ion energy equations together and neglecting terms of the order $(m_e/m_p)^{1/2}$ and higher gives:
\bea
\label{eq:energy_i}
 \frac{\partial}{\partial t}\left(\ve_i + \frac{P_e}{\gamma-1}\right) &  +& 
 \nabla\cdot\left(\ve_i \mb{v}_i + \frac{P_e\mb{v}_e}{\gamma-1} +
  \mb{v}_i\cdot\mathbb{P}_i + \mb{v}_e\cdot\mathbb{P}_e + \mb{h}_i +
  \mb{h}_e\right) =  \mb{j}\cdot\mb{E}
\nonumber \\ 
&+& \mb{v}_i\cdot\mb{R}_i^{in}  + Q_i^{in}
- \Gamma_n^{rec}\frac{1}{2} m_i v_i^2 - Q_{n}^{rec}
+ \Gamma_i^{ion}(\frac{1}{2} m_i v_n^2 - \phi_{ion} )
+ Q_i^{ion}  \nonumber \\
& + &
\Gamma^{cx}\frac{1}{2}m_i\left(v_{n}^{2} - v_{i}^{2}\right) 
+ \mb{v}_n \cdot \mb{R}_{in}^{cx} - \mb{v}_i \cdot \mb{R}_{ni}^{cx}
+ Q_{in}^{cx} - Q_{ni}^{cx}.
\eea 
The neutral energy equation is
\bea
\label{eq:energy_n}
\frac{\partial \ve_n}{\partial t} &+& \nabla \cdot (\ve_n \mb{v}_n +
\mb{v}_n \cdot \mathbb{P}_n + \mb{h}_n)\nonumber \\
& = &   -\mb{v}_n \cdot \mb{R}_i^{in} +
 Q_n^{ni}  - \Gamma_i^{ion}\frac{1}{2} m_i\mb{v}_n^2 -
Q_i^{ion}
+ \Gamma_n^{rec}\frac{1}{2} m_i\mb{v}_i^2 + Q_n^{rec} \nonumber \\
&+& \Gamma^{cx} \frac{1}{2} m_i (\mb{v}_i^2 - \mb{v}_n^2) + \mb{v}_i
\cdot \mb{R}_{ni}^{cx} - \mb{v}_n\cdot\mb{R}_{in}^{cx} + Q_{ni}^{cx} -
Q_{in}^{cx}. \eea 
\nind Here $\ve_\alpha\equiv m_\alpha n_\alpha v_\alpha^2/2 +
P_\alpha /(\gamma-1)$ is the internal energy density of 
fluid $\alpha$, and the term $\Gamma_i^{ion}\phi_{ion}$ represents optically
thin radiative losses. The ratio of specific heats is denoted by $\gamma$.
 $Q_\alpha^{\alpha\beta}$ is the heating of species $\alpha$
 due to interaction with species $\beta$, which is a combination of frictional
 heating and a thermal transfer between the two populations:
$Q_\alpha^{\alpha\beta} = \mb{R}_\alpha^{\alpha\beta}\cdot(\mb{v}_{\beta} - \mb{v}_{\alpha}) + m_{\alpha\beta}n_{\alpha}\nu_{\alpha\beta}(T_{\beta}-T_{\alpha})$.
 The heat
fluxes $\mb{h}_e$, $\mb{h}_{i}$ are calculated using the Braginskii closure for a magnetized
plasma, and can be written as 
\be
\mb{h}_{\alpha} = \left[ \kappa_{\|,\alpha}\hat{\mb{b}}\hat{\mb{b}} + \kappa_{\bot,\alpha}(\mathbb{I}-\hat{\mb{b}}\hat{\mb{b}}) \right]\cdot\nabla k_{B} T_\alpha \ee
where $\kappa_{\|,\alpha}(T_\alpha)$ and $\kappa_{\bot,\alpha}(n_\alpha,T_\alpha,|\mb{B}|)$ account for the effects of thermal diffusion parallel to and perpendicular to the magnetic field direction ($\hat{\mb{b}}$) respectively, and whose functional forms can be found in \citet{1965RvPP....1..205B}.

The neutral thermal conduction is isotropic $\mb{h}_{n} = -\kappa_{n}\nabla k_{B} T_{n}$.
$Q_{\alpha}^{r}$ denotes thermal energy gain of species $\alpha$ due
to a reaction \textit{r}, with $Q_{i}^{ion} = \Gamma_{i}^{ion}\frac{3}{2}k_{B}T_{n}$ and $Q_{n}^{rec} = \Gamma_{n}^{rec}\frac{3}{2}k_{B}T_{i}$.
$Q_{\alpha\beta}^{cx}$ denotes heat flow
from species $\beta$ to species $\alpha$ due to charge exchange \citep{Meier11,Meier12b}. The
temperature of the neutrals is given by $T_{n} = P_{n}/(n_{n}k_{B})$;
and the ion and electron temperatures are assumed to be equal
such that $T_i = P_{i}/(n_{i}k_{B}) = P_{e}/(n_{e}k_{B}) = T_e$.

\nind\textit{Ohm's Law:} \\
The generalized Ohm's law is given by the electron momentum equation:
\bea
\label{eq:mom_e}
\frac{\partial}{\partial t} (m_e n_e \mb{v}_e) + 
\nabla \cdot (m_e n_e \mb{v}_e \mb{v}_e) 
& - & \Gamma_n^{ion} m_e \mb{v}_n + \Gamma_i^{rec} m_e \mb{v}_e \nonumber \\
& = &- e n_e (\mb{E} + \mb{v}_e \times \mb{B}) -
\nabla\cdot\mathbb{P}_e + \mb{R}_{e}^{ei} + \mb{R}_{e}^{en}.
\label{eqn:ohms1}
\eea

In the HiFi implementation all terms on the left hand side of
Equation (\ref{eqn:ohms1}) are neglected, as in the chromosphere of the Sun, 
the electron inertial scale $c/\omega_{pe}$ is likely much smaller than magnetic diffusion length scales.
  However, the electron viscous stress tensor is preserved in
order to represent the effects of microturbulence and 3D instabilities \citep{2011Natur.474..184C}, and also to 
damp the dispersive Whistler and kinetic Alfv\'{e}n waves at
the shortest resolvable wavelengths. Note that the electron-neutral
collision term $\mathbf{R}_{e}^{en}$ cannot be neglected. Using
the identity $\mb{v}_{i}=\mb{v}_{e}+\mb{j}/en_{i}$, and the definition
$\mathbf{w}\equiv\mathbf{v}_{i}-\mathbf{v}_{n}$, Eq.~(\ref{eqn:ohms1})
can then be written as 
\be \mb{E} + (\mb{v}_i \times \mb{B}) =
\eta\mathbf{j} + \frac{\mb{j}\times\mathbf{B}}{en_{i}}
-\frac{1}{en_{i}}\nabla\cdot \mathbb{P}_{e} -
\frac{m_{e}\nu_{en}}{e}\mathbf{w},
\label{eqn:ohms}
\ee
where 
\be
\eta=\frac{m_{e}n_{e}(\nu_{ei}+\nu_{en})}{(e n_{e})^2}
\ee is the electron, or Spitzer, resistivity, which includes electron-ion and electron-neutral collisions. 
The system is closed by the use of Faraday's law $\frac{\partial B}{\partial t}  = -\nabla\times\mb{E}$.

It is worth noting how the model presented here differs from the partially ionized single-fluid model used in 
\citet{2006A&A...450..805L} and \citet{2007ApJ...666..541A}.
 The single-fluid approach implicitly assumes that the ions and neutrals are in ionization balance, which is determined by the average 
density and temperature, and the center of mass velocity  used in the Ohm's law is an average over ions and neutrals. The Pedersen resistivity, present in this partially ionized single-fluid approach, is a consequence of this center of mass velocity. The multi-fluid model presented here follows the ions and neutrals separately, thus self-consistently including the interactions between ions and neutrals which are represented by the Pedersen resistivity in the single-fluid approach. A key advantage of the multi-fluid model used in this paper is that ions and electrons are allowed to be out of local thermodynamic equilibrium (LTE). This will be shown to be vital for  the onset of fast reconnection in chromospheric plasmas.


\subsection{Normalization}
The equations are non-dimensionalized by dividing each variable ($C$) by its normalizing value ($C_{0}$).
The set of equations requires a choice of three normalizing values. Normalizing values for the length ($L_{0}=1\times10^{5} ~ \textrm{m}$), number density ($n_{0}=3.3\times10^{16} ~ \textrm{m}^{-3}$), and magnetic field ($B_{0}=1\times10^{-3} ~ \textrm{T} $) are chosen. From these values the normalizing values for the velocity ($v_{0}=B_{0}/\sqrt{\mu_{0}m_{p}n_{0}}=1.20\times10^{5} ~ \textrm{m}/\textrm{s}^{-1}$), time ($t_{0}=L_{0}/v_{0} =0.83$ s), temperature ($T_{0}=m_{p}B_{0}^{2}/k_{B}\mu_{0}m_{p}n_{0} = 1.75\times10^{6}~\textrm{K}$), pressure ($P_{0}=B_{0}^2/\mu_{0}=0.80 ~ \textrm{Pa}$) and resistivity ($\eta_{0} = \mu_{0}L_{0}v_{0} = 1.5\times10^{4} ~ \Omega$m), can be derived. 

\subsection{Initial conditions and simplified equations}

The simulation domain extends from -36$L_{0}$ to 36$L_{0}$ in the $x$ direction and -6$L_{0}$ to 6$L_{0}$ in the $y$ direction, with a periodic boundary condition in the $x$-direction and perfectly-conducting boundary conditions in the $y$-direction.  The size of the domain has been chosen so that the boundaries and the particular boundary conditions do not affect any properties of the reconnection region centered and localized near $x=0, ~ y=0$. 

The initial neutral fluid number density is 200$n_{0}$ ($6.6\times 10^{18} ~ \textrm{m}^{-3}$), and the ion fluid number density is $n_{0}$ ($3.3 
\times 10^{16} ~ \textrm{m}^{-3}$). This gives a total (ion+neutral) mass
density of $1.11\times10^{-8} ~ \textrm{kg}.\textrm{m}^{-3}$, and an
initial ionization level ($\psi_{i}\equiv n_{i}/(n_{i}+n_{n}$)) of 0.5
\%. The initial electron, ion and neutral temperatures are set to 0.005$T_{0}$ ($8750 \textrm{K})$. These initial conditions are consistent with lower to middle chromospheric conditions, based on 1D semi-empirical models of the quiet Sun \citep{1981ApJS...45..635V}.\\

In this plasma parameter regime, ion-neutral identity preserving
collisions and charge exchange (CX) interactions are equally important
and have a very similar effect of collisionally coupling neutral and
ion fluids with a neutral-ion collision mean free path of $L_{ni}=v_{T,n}/\nu_{n,i}=140 ~ \textrm{m}$. 
  However, the detailed CX physics is substantially
more complicated and so for simplicity of interpretation CX interactions
are neglected in this initial study and will be considered in
future work.  Charge exchange terms (terms with the superscript $cx$)
in Equations (\ref{eq:mom_i})-(\ref{eq:energy_n}) are thus dropped.

The ion inertial scale for these plasma parameters
is $c/\omega_{pi}\approx 1 m$, which is much smaller than any scale
of interest. Consequently, we neglect the Hall ($\mb{j}\times\mb{B}$)
and the electron pressure tensor ($\nabla\cdot\mbb{P}_{e})$ terms in
Equation (\ref{eqn:ohms}), the electron viscous stress tensor
($\nabla\cdot\pi_{e})$ in Equation (\ref{eq:mom_i}) and
Equation (\ref{eq:energy_i}), and set electron velocity $\mb{v}_e$ equal to
ion velocity $\mb{v}_i$ in Equation (\ref{eq:energy_i}).  Similarly,
\citet{2011ApJ...739...72M} showed that in the geometry of a
reconnection current sheet electron-neutral collisions are only
important in calculating resistivity, and so the term 
$\frac{m_{e}\nu_{en}}{e}\mathbf{w}$ in Equation (\ref{eqn:ohms}) is also dropped.
With these simplifications, we solve the following set of governing PDEs:

\nind\textit{Continuity: } \\
\bea
 \frac{\partial n_i}{\partial t} + \nabla \cdot (n_i \mb{v}_i) & =
& \Gamma_i^{ion} + \Gamma_i^{rec}, \\
 \frac{\partial n_n}{\partial t} + \nabla \cdot (n_n \mb{v}_n) & 
= & \Gamma_n^{rec} + \Gamma_n^{ion}.
 \eea
\nind\textit{Momentum: } \\
\bea
\frac{\partial}{\partial t} (m_i n_i \mb{v}_i) + 
\nabla \cdot (m_i n_i \mb{v}_i \mb{v}_i + \mathbb{P}_i+ P_e) & = &
 \mb{j} \times \mb{B} + \mb{R}_i^{in} 
+ \Gamma_i^{ion} m_i \mb{v}_n -
\Gamma_n^{rec} m_i \mb{v}_i, \\
\frac{\partial}{\partial t} (m_i n_n \mb{v}_n) +  \nabla \cdot (m_i
n_n \mb{v}_n \mb{v}_n + \mathbb{P}_n) & = & - \mb{R}_i^{in} 
+  \Gamma_n^{rec} m_i \mb{v}_i
-\Gamma_i^{ion} m_i \mb{v}_n . 
\eea 
\nind\textit{Internal Energy:} \\
\bea
 \frac{\partial}{\partial t}\left(\ve_i + \frac{P_e}{\gamma-1}\right) & + &
 \nabla\cdot\left(\ve_i \mb{v}_i + \frac{\gamma P_e\mb{v}_i}{\gamma-1}  +
  \mb{v}_i\cdot\mathbb{P}_i + \mb{h}_i +
  \mb{h}_e\right) \\
  & = &   \mb{j}\cdot\mb{E} 
  +  \mb{v}_i\cdot\mb{R}_i^{in}  + Q_i^{in}
- \Gamma_n^{rec}\frac{1}{2} m_i v_i^2 - Q_{n}^{rec} \\
& +& \Gamma_i^{ion}(\frac{1}{2} m_i v_n^2 - \phi_{ion} )
+ Q_i^{ion}, \\
\frac{\partial \ve_n}{\partial t} &+& \nabla \cdot (\ve_n \mb{v}_n 
\mb{v}_n \cdot \mathbb{P}_n + \mb{h}_n)
=  -\mb{v}_n \cdot \mb{R}_i^{in} + Q_n^{ni} \nonumber \\
&-& \Gamma_i^{ion}\frac{1}{2} m_i\mb{v}_n^2 - Q_i^{ion}
+ \Gamma_n^{rec}\frac{1}{2} m_i\mb{v}_i^2 + Q_n^{rec}.
 \eea 
\nind{\textit{Ohm's Law:}\\
\be
\mb{E} + (\mb{v}_i \times \mb{B}) = \eta\mathbf{j}.
\ee
 Note that to investigate the scaling of the reconnection rate with the Lundquist number (S), the Spitzer resistivity $\eta$ is made a parameter of the simulations. The values used for $\eta$ are $[0.5, 1,2,4,8,20]\times 10^{-5} \eta_{0}$, and so $\eta$ lies in the range [0.08,3] $\Omega$m. 

The viscosity coefficients for neutrals and ions are set to $\xi_{i} = \xi_{n} = 10^{-3}\xi_{0}$ and the neutral thermal conduction coefficient is $\kappa_{n
}=4\times10^{-3}\kappa_{0}$. The normalizing constants are $\xi_{0} = m_{p}n_{0}L_{0}v_{0}$ and $\kappa_{0}=m_{p}n_{0}L_{0}v_{0}^3/T_{0}$. Note that for these plasma parameters the isotropic neutral heat conduction is much faster than any of the anisotropic heat conduction tensor components for either ions or electrons, with collisional ion-neutral heat exchange $Q_{i}^{in} \gg \nabla\cdot(\mathbf{h}_i + \mathbf{h}_e)$.  Thus, thermal diffusion for all species is dominated by neutral heat conduction and is primarily isotropic.

A Harris current sheet is used for the initial magnetic configuration, and is given here in terms of the in-plane magnetic flux $A_{z}$:
\be
A_{z} = - B_{0}\lambda_{\psi}\ln{\cosh{(y/\lambda_{\psi})}}
\ee
where $\mathbf{B} = \nabla \times A_{z}\hat{\mathbf{e}}_{z}$ and $\lambda_{\psi}=0.5L_{0}$ is the initial width of the current sheet.

To provide the outward force to maintain this current sheet, both the
ionized pressure $P_{p} =P_{i}+P_{e}$ and the neutral pressure
$P_{n}$ are increased in the sheet:
\bea
P_{p}(y) & = & P_{p} + \frac{1}{2}\frac{F}{\cosh^{2}{(y/\lambda_{\psi})}}, \\
P_{n}(y) & = & P_{n} + \frac{1}{2}\frac{1-F}{\cosh^2{(y/\lambda_{\psi})}},
\eea
where $F=0.01P_{0}$ is chosen to maintain an approximate ionization
balance of 0.5\% inside the current sheet. These perturbations ensure that the total (ion and neutral) pressure perturbation balances the Lorentz force of the magnetic field in the current sheet:
\be
0 =  -\nabla (P_{n}+P_{p}) + \mathbf{j}\times\mathbf{B}.
\ee
At the same time,  a relative velocity between
ions and neutrals is required, so that the frictional force
$\mb{R}_{i}^{in}$ can couple the ionized and neutral fluids and keep the fluids individually in  approximate force balance:
\begin{eqnarray}
0 & = & -\nabla P_{p} + \mathbf{j}\times\mathbf{B} + \mb{R}_{i}^{in}, ~ \textrm{and} \\
0 & = & -\nabla P_{n} - \mb{R}_i^{in}.
\end{eqnarray}
 To give this balance, we choose an initial ion velocity
\be
{v_{i}}_{y}(y) = \frac{(F-1)}{n_i n_n\nu_{in}}\frac{\tanh{(y/\lambda_{\psi})}} {\lambda_{\psi}\cosh^{2}{(y/\lambda_{\psi})}}\frac{n_{0}^2v_{0}^2}{P_{0}}.
\ee

To this initial steady state a small, local perturbation of the flux, $A_{z}^{1}$ is added to initiate the primary tearing instability and start the magnetic reconnection:
\be
A_{z}^{1}= - \epsilon e^{-{\left(\frac{x}{4\lambda_{\psi}}\right)}^{2}}e^{-{\left(\frac{y}{\lambda_{\psi}}\right)}^{2}}
\ee
where $\epsilon = 0.01B_{0}L_{0}$.
The initial state is shown in Figure \ref{fig:IC1}. Only a subset of the domain is shown, and the $y$ axis is stretched by a factor of 40, so that thin structures can be clearly seen in the plots. Using the symmetry of the initial conditions, only the top right quadrant of the domain is simulated, with the use of appropriate boundary conditions.

\section{Results}

\subsection{Decoupling of inflow during magnetic reconnection}

As mentioned in the previous section, $\eta$ is made a parameter of the simulations, and is not dependent on the local plasma parameters, so that a general scaling law of reconnection rate with resistivity can be derived.
 Six simulations are performed which have values of the resistivity of $[0.5, 1,2,4,8,20]\times 10^{-5} \eta_{0}$. Firstly, the generic processes that are evident in the simulations are highlighted by focusing on one particular value of $\eta=0.5\times10^{-5}\eta_{0}$. Later, a relationship between reconnection rate and $\eta$ is determined for the range of $\eta$ in these simulations. 

Figure \ref{fig:reconnection} shows the early evolution of the
reconnection region resulting from the initial perturbation. The
Harris current sheet undergoes the tearing instability, magnetic field
reconnects at the X-point and flux is ejected. By $t=537.5 t_{0}$ a
Sweet-Parker-like reconnection region has been formed. At this stage, ions are pulled in by the magnetic field and drag the neutrals with them.

Figure \ref{fig:decoupling} shows the ion and neutral flow fields, on
a smaller subdomain of the simulation, centered on the reconnection region. Panel \ref{fig:decoupling}a) shows the vertical velocities at time $t=537.5 t_{0}$,  with ions on the top left quadrant and neutrals on the top right  quadrant.
Panel \ref{fig:decoupling}b) shows the magnitude of the horizontal
velocity, again with  ions on the top left quadrant and neutrals on
the top right  quadrant. The bottom two quadrants of panels a) and b)
show the flow vectors for ions, on the left, and for neutrals, on the
right, both scaled to the same magnitude. The vertical velocities in
panel \ref{fig:decoupling}a) show that the ions and neutrals are
decoupled, with the ions flowing faster into the reconnection region than the neutrals. The difference between neutral and ion velocity is approximately 90\% of the ion velocity.
The horizontal velocities in panel \ref{fig:decoupling}b) show that the ion and neutral outflows (the strong horizontal velocity at  x=+/-2.5$L_{0}$) are coupled. The difference between the ion and neutral outflow is negligible compared to the actual flow.
 The Alfv\'{e}n speed can be estimated using the upstream magnetic field of $B_{up}=0.6B_{0}$, taken at the point in the current sheet where the current amplitude reaches half its maximum value, and the total number density at the center of the sheet of 280$n_{0}$ to  give $v_{A} = 0.035v_{0}$. The coupled outflow of ions and neutrals, which has a maximum of $0.015v_{0}$, is thus approximately half the Alfv\'{e}n speed (based on total number density, as the outflow is coupled) at this time. Later in time the outflow increases to the Alfv\'{e}n speed.

This feature of the $\eta=5\times10^{-6}\eta_{0}$ simulation is common
to all six simulations. The ions and neutrals inflows are decoupled,
as ions are pulled in by reconnecting magnetic field and the
neutrals are dragged in via collisions. The timescale of inflow is fast enough that
the collisions cannot keep the neutrals completely coupled to the ions
and an excess of ions builds up in the reconnection region, creating an 
ionization imbalance. This is the situation considered for astrophysical plasmas by both \citet{1999ApJ...511..193V}, who treated weakly ionized reconnection with an analytic approach, and \citet{2003ApJ...583..229H}, who calculated analytic and numerical solutions of 1D steady state models of weakly ionized reconnection.

\subsection{Reconnection rate scaling with resistivity}
 
As described in the previous section, the plasma in the reconnection region is out of ionization balance as the neutrals are largely left behind by the ions.
 Figure \ref{fig:steady_state} shows the four components contributing
 to $\frac{\partial n_i}{\partial t}$ in the ion continuity
 equation. The top left quadrant shows the loss due to recombination,
 the bottom left quadrant shows the  loss due to outflow (horizontal
 gradient in horizontal momentum of ions), the  top right quadrant
 shows the gain due to inflow (vertical gradient in vertical
 momentum), and the bottom right shows the gain due to ionization. The
 color scheme is based on a log-scale, and shows that within the reconnection
 region gains due to ionization are negligible relative to the other
 three terms. Looking at the largest values for the remaining three
 terms, the losses due to recombination and outflow are comparable,
 and add up to equal the gain due to inflow. This shows that the
 reconnection region is close to a steady state, with inflow of ions balanced by comparable contributions from recombination and outflow.  Recall that the 1D models of  \citet{1999ApJ...511..193V} and \citet{2003ApJ...583..229H} assumed that recombination was fast enough to dominate over the outflow, and thus the horizontal direction (along the current sheet) was ignorable.  Instead, Figure \ref{fig:steady_state} shows that for the self-consistently created reconnection region in this parameter range the recombination and outflow are comparable and so 2D effects cannot be neglected.
 
 The nature of the steady state balance in these reconnection simulations has important consequences for the scaling of the reconnection rate. In the standard Sweet-Parker reconnection scenario, inflow of ions into the reconnection region is assumed to be balanced by outflow of ions. This, along with other assumptions, gives the standard scaling of the normalized inflow rate (reconnection rate) of $ M\propto \sqrt{\eta} \propto 1/\sqrt{S}$. For the weakly ionized plasma in these simulations, the situation is very different due to the ionization imbalance.

The standard steady-state argument of the Sweet-Parker model can be modified to include the reacting multi-fluid equations, in order to derive a scaling relationship for the reconnection rate. Figure \ref{fig:SP_PIP} shows a cartoon of the reconnection region in these simulations. The reconnection region, inside which the frozen-in constraint is broken by resistivity $\eta$, has width $\delta$ and length $L$. There is an external ion density of $n_{i,ext}$ and a current sheet ion density of $n_{i,CS}$. The inflow of ions into the reconnection region is $v_{in}$, the outflow of ions  is $v_{out}$, and the upstream magnetic field is $B_{up}$. Assuming that the system is in steady state, i.e., that $\frac{\partial n_{i}}{\partial t}=0$, and integrating around the reconnection region gives
\be
n_{i,ext}v_{in}L = n_{i,CS}(\delta v_{out} + \delta L \nu^{rec} - \delta L \nu^{ion}),
\ee
where $\nu^{rec}$ and $\nu^{ion}$ are the recombination and ionization frequencies, defined in Equations (\ref{eqn:recomb}) and (\ref{eqn:ioniz}). Defining $\nu_{out} \equiv v_{out}/L$, and 
equating  Ohm's law evaluated inside (${E}=\eta{j}$) and outside (${E}=v_{in}B_{up}$) of the reconnection region,
\be
v_{in}B_{up} = \eta j \approx \frac{\eta B_{up}}{\delta \mu_{0}},
\ee
to eliminate $\delta$, this steady state equation can be rewritten as
\be
v_{in} \approx \sqrt{ \frac{\eta}{\mu_{0}}\frac{n_{i,CS}}{n_{i,ext}} (\nu_{out} + \nu^{rec} - \nu^{ion})}.
\label{eqn:rates}
\ee

For a plasma in ionization balance,  $\nu^{rec} = \nu^{ion}$, and $n_{i,ext} \approx n_{i,CS}$. Using these relationships  and the total momentum equation to derive $v_{out} = v_{A}$ (where $v_{A} = \frac{B_{up}}{\sqrt{\mu_{0}\rho_{total}}}$), recovers the standard Sweet Parker scaling law
\be
M\equiv \frac{v_{in}}{v_{A}} \approx \sqrt{\frac{\eta}{v_{A}L\mu_{0}}} = \sqrt{\frac{1}{S}},
\ee
where $B_{up}$ is the magnetic field upstream of the current sheet,
and $\rho_{total}$ is the total (ion + neutral) density in the
reconnection region.

In contrast, in these reacting two-fluid simulations, the system non-linearly and self-consistently forms a current layer where plasma is out of ionization balance, where the recombination is comparable to the outflow, and where ionization is negligible compared to both recombination and outflow. Hence the inflow rate can be approximated by
\be
v_{in} \approx \sqrt{ \frac{\eta}{\mu_{0}}\frac{n_{i,CS}}{n_{i,ext}}(\nu_{out} + \nu^{rec}) }.
\ee
Note that this equation does not by itself indicate how the reconnection rate will depend on $S$, as it is not clear how $\nu^{rec},n_{i,CS}$, and $n_{i,ext}$ depend on $S$ from the equations. We will therefore use our series of simulations performed over a range of $\eta$ to determine the reconnection rate dependence on $S$. 

For all six simulations the effective Lundquist number is defined by 
\be
S_{sim} \equiv \frac{v_{A,0}L_{sim}\mu_{0}}{\eta},
\ee
where $L_{sim}=2.75L_{0}$. Note that $L_{sim}$ is much smaller than the horizontal extent of the domain.  We define $v_{A,0}$ as the Alfv\'{e}n velocity given a field strength of $B_{0}$ and number density $201n_{0}$, i.e., based on the initial background magnetic field and number density. Note that $S_{sim}$ varies over simulations due to $\eta$ only.
The reconnection rate is defined by 
\be
M_{sim} \equiv \frac{\eta j_{max}}{v_{out}B_{up}}.
\ee
Here, $j_{max}$ is the maximum value of the current density, located at 
 $(x,y)=(x_{j},0)$, within the
reconnection region, and $B_{up}$ is 
evaluated at $(x_{j},\delta_{sim})$ where $\delta_{sim}$ is the $y$
location on the line $x=x_{j}$ at which the current density reaches
$j_{max}/2$, i.e., half-width at half-max of the current sheet.


For each simulation, the value of $M_{sim}$ is taken at a time when the 
length of the current sheet, defined by the distance from the original X-point 
to the location of maximum outflow $v_{out}$ equals $L_{sim}$. This happens at a different time
for each simulation, but in all cases prior to the onset of any secondary
instabilities of the current sheet.
This instantaneous value of $M_{sim}$ is plotted for each simulation against the effective 
Lundquist number $S_{sim}$ in Figure \ref{fig:rec_rate}. 
The dashed line shows the single-fluid Sweet-Parker predicted
scaling $M \propto 1/\sqrt{S}$. Over the range of $S$ which these six simulations cover, the
reconnection rate is only weakly dependent on the Lundquist number, with a slow decrease in $M_{sim}$
with increasing $S_{sim}$. The reconnection rate is 
much faster than the Sweet-Parker reconnection rate, as a direct result
of the decoupling of ions from neutrals and the enhanced recombination
in the reconnection region. As discussed in section \ref{secondary}, plasmoid formation increases the reconnection rate for $S_{sim} > S_{c}$. We also note that each of these simulations have been performed with the same viscosity coefficient, and it is known that the reconnection rate dependency on resistivity weakens when viscous effects in the reconnection layer become important \citep{1984PhFl...27..137P}.

Figure \ref{fig:width} shows the scaling of the reconnection region aspect ratio $\sigma_{sim}\equiv (\delta_{sim}/L_{sim})$ 
 with $S_{sim}$ for the six simulations. Also shown is the Sweet-Parker scaling of $1/\sqrt{S}$. The simulations do not exhibit the $\sigma \propto 1/\sqrt{S}$ scaling but show
 an approximate $1/S$ scaling (the power law fit through the
 simulation points has an exponent of $-1.1 \pm 0.17$).  This $1/S$ scaling is consistent with the result of Figure \ref{fig:rec_rate}, that $M_{sim}$ is approximately independent of $S_{sim}$. This is shown as follows: Assume that $v_{out}$ does not substantially vary over different simulations, and note that $S_{sim}$ only changes over simulations due to $\eta$. Then using the definition of $M_{sim}$ and $j_{max} \approx B_{up}/\delta_{sim}$ gives $M_{sim} \propto \eta j_{max}/B_{up} \approx \eta/\delta_{sim} \propto 1/S_{sim}\sigma_{sim} \approx \textrm{const.}$

\subsection{The secondary (plasmoid) instability}
\label{secondary}
As discussed above, at a critical aspect ratio of $\sigma_{c}=1/200$, a resistive current sheet can become unstable to a secondary tearing instability known as the \textit{plasmoid instability}. 
Of the six two-fluid simulations in this paper, three cases show evidence of secondary tearing. 
Figure \ref{fig:plasmoid} shows the onset of the plasmoid instability for the simulation with $\eta=0.5\times10^{-5}\eta_{0}$, with the two contours of magnetic flux at $A_{z}=-0.0681B_{0}L_{0}$ (white line) and $A_{z}=-0.0687B_{0}L_{0}$ (black line) following the evolution of two particular field-lines in time. The initial laminar current sheet breaks up into a number of plasmoids, with thinner current sheets between them. 
The current density and recombination rate within this fragmented reconnection region are shown in Fig.~\ref{fig:plasmoid} on a pseudo-color scale.  During this phase of the 
 reconnection the formation of a plasmoid chain redistributes the ionized
  plasma within the current sheet, which is otherwise approximately
  uniformly distributed throughout the reconnection region, into regions of higher and lower electron number density. Since $\Gamma_{i}^{rec}\propto n_e^2$, the recombination rate is increased within the plasmoids with respect to the sub-layers between them. This leads to the plasmoid magnetic flux collapsing on itself (i.e,  disappearing as it would in vacuum) at a rate comparable to or faster than the plasmoids are exhausted out of the reconnection region. This collapse can affect the expected distribution of plasmoid sizes and the ionization fraction within the plasmoids relative to the background medium in the reconnection exhaust.
  
\citet{Huang10} found that in a fully ionized plasma, the onset of the plasmoid instability occurred at a 
critical value of Lundquist number of $4\times10^{4}$ which, with the Sweet-Parker scaling of current sheet width with Lundquist 
number, corresponds to a critical current sheet aspect ratio of $1/200$, as shown by the dashed line in Figure 
 \ref{fig:width}. 
The three simulations that undergo secondary tearing are shown as
diamonds in Figure \ref{fig:width}. The intersection of a power law fit to the $\sigma_{sim}(S_{sim})$ data
intersects the $\sigma_{sim}=1/200$ line at a value of $S_{sim}=10^{4}$.
The simulation with $S_{sim}=10^4$ exhibits the plasmoid instability, as do the two simulations 
with higher $S_{sim}$ (shown as diamonds in the Figure). 
These two facts support the
postulation of \citet{Loureiro07} that the criterion for the plasmoid instability is that the aspect ratio decreases below a critical value.
For our simulations, as for \citet{Huang10}, this critical value is 1/200.

To demonstrate the change in reconnection rate due to the plasmoid instability, 
$M_{sim}$ is evaluated again later in each of the three plasmoid-unstable simulations.
It is difficult to measure $M_{sim}$ during the plasmoid instability, as the reconnection rate varies with time, depending
on the plasmoid evolution. To give an idea of how the plasmoid instability is affecting the reconnection rate
we plot in red in Figure \ref{fig:rec_rate}  the range of $M_{sim}$ observed after plasmoid formation in each simulation 
until the current sheet width becomes spatially unresolved.
It is apparent, particularly in higher $S_{sim}$ simulations,
that the reconnection rate increases as the plasmoids
develop. Thus, the reconnection rate in this regime is determined
by both the fast recombination of ions, and the effect of the
secondary tearing instability.

\section{Discussion}

In this paper magnetic reconnection in the solar chromosphere is simulated using a partially ionized reacting multi-fluid plasma model. The number densities and ionization levels are consistent with lower to middle chromospheric conditions. The dependence of reconnection rate on Lundquist number is investigated by setting the resistivity to be a parameter of the simulations.
A simple 2D Harris current sheet configuration with a local
perturbation to the in-plane flux is used. The system
self-consistently and non-linearly creates a reconnection region which
is out of ionization balance, due to the decoupling of ion and neutral
inflows. The model is able to capture this physical effect as the two
fluids are followed separately, and the ionization and recombination
rates are self-consistently calculated based on local plasma
parameters. In the reconnection region, recombination of excess ions is comparable to the outflow of ions, which leads to fast reconnection independent of the Lundquist number, and when normalized properly, the reconnection rate is approximately 0.1.
  It is worth noting that guide (out of plane) magnetic field, not included in these simulations, will have an effect on the scaling, as the flux associated with the guide field can inhibit fast reconnection \citep{2003ApJ...590..291H}.

The onset of fast reconnection in weakly ionized astrophysical plasmas was predicted by \citet{1999ApJ...511..193V} and \citet{2003ApJ...583..229H}, assuming that recombination dominates outflow in the current sheet so the system could be treated with one dimensional analysis. In this regime \citet{2003ApJ...583..229H} found that fast reconnection occurred when
\be
Z=\frac{\beta_0^{3/\gamma}}{10}\frac{t_{recomb}t_{\Omega}}{t_{AD}^2} < Z_{c}
\ee
where $Z_{c}$ was either $10^{-4}$ for the numerical solution, or $10^{-2}$ for the theoretical solution, 
$\beta_0\equiv\frac{P_i+P_e}{B^2/\mu_0}$, $t_{rec}$ is the timescale of recombination,  $t_{\Omega}  \equiv  L^2/\eta$, and $t_{AD} \equiv L^2/\eta_{p}$. In  our multi-fluid simulations, the smallest resistivity is $\eta \approx 0.08 ~ \Omega \textrm{m}$, $\eta_{p} \approx 0.3 ~ \Omega \textrm{m}$, and $t_{rec} =1/\nu^{rec} \approx 30 ~ \textrm{s}$, and so $Z\approx4\times10^{-5}$, which lies in the `fast' reconnection regime of 
\citet{2003ApJ...583..229H}. While we have shown that while the prediction of fast reconnection by \citet{1999ApJ...511..193V} and \citet{2003ApJ...583..229H} holds in the chromosphere, in 2D reconnection  the recombination is not the dominant mechanism for the removal of ions from the reconnection region, as outflows are equally important.

The critical aspect ratio at which secondary tearing sets in, $\sigma_{c}=1/200$, is reached at a lower Lundquist number than has previously been seen in single-fluid simulations with $\beta \approx 1$ \citep{Huang10,Samtaney09}. This is because in this fast reconnection regime $\sigma \propto 1/{S}$, compared to Sweet-Parker reconnection where
$\sigma \propto 1/\sqrt{S}$. Note that \citet{2012PhPl...19g2902N} found that for $\beta = 50$, the onset occurs at a Lundquist number of 2000-3000, and an aspect ratio of 1/60. The plasmoids which form due to secondary tearing in these simulations are losing ions due to recombination, and so their evolution is potentially very different from those seen in fully ionized simulations. The further evolution of the plasmoid  beyond the initial formation and collapse seen here is left to a follow-up investigation. 

It has been conjectured that current sheets are ubiquitous in the chromosphere \citep{2012ApJ...751...75G}. It has also been established that spicules and chromospheric jets are ubiquitous solar phenomena \citep{2000SoPh..196...79S}. Magnetic reconnection in chromospheric current sheets is therefore a promising mechanism to link these two ubiquitous phenomena, and would explain the formation of spicules and chromospheric jets. These simulations show that the chromosphere can exhibit fast reconnection with rates that are comparable to the estimated reconnection rates of observed outflows, as well as the creation of plasmoids, or ``blobs" of plasma, due to the secondary tearing mode. The reconnection outflow velocities we find in these simulations are 5 km/s, which is close to the speeds observed for chromospheric jets, but is small compared to the speeds of 20-150 km/s observed for spicules. Using the scaling argument that $v_{out} \propto B_{up} /\sqrt{n_i+n_n}$, we expect that by increasing the magnetic field from 10G to 50G, and by moving higher up in the chromosphere where the neutral density is an order of magnitude lower (and the ionization level increases from 0.5\% to 10\%) spicule-like outflow velocities of up to 80 km/s can realistically be achieved.

 The current sheets formed in these simulations of chromospheric reconnection are out of ionization balance to such a degree that short recombination times of 30 s are created. These
times are much smaller than the estimated 3 minute  recombination time of an acoustic shock heated chromosphere \citep{2002ApJ...572..626C}, highlighting
magnetic reconnection as the prime transient phenomena in the chromosphere driving non-LTE recombination.
The major advantage of the multi-fluid approach over single-fluid models is that it can self consistently produce non-LTE high ion density structures.
This paper has shown that such ion density structures form in magnetic reconnection, and that the resulting fast recombination affects the reconnection physics. Hence multi-fluid simulations such as these 
are vital to understanding transient phenomena in the chromosphere.

Finally, it should be noted that \citet{2004PhPl...11.2199S} and
others have argued that the addition of the ion inertial and/or finite
Larmor radius effects to single-fluid resistive MHD is essential to
obtain the resistivity-independent ``universal'' fast reconnection rate
of $M\approx 0.1$. The simulation results presented here yield this
same resistivity-independent reconnection rate without any of the
ion-electron decoupling effects; relying instead on the combination of
enhanced recombination rate with generation of the secondary plasmoids
in the reconnection region.

\begin{acknowledgments}
\nind{Acknowledgements:}
This work has been supported by the NASA Living With a Star \& Solar and Heliospheric 
Physics programs, the ONR 6.1 Program, the U.S. DOE Experimental
Plasma Research program, by LLNL under Contract DE-AC52-07NA27344, and by the NRL-\textit{Hinode} analysis program.
The simulations were performed under a grant of computer time from the DoD HPC program.
\end{acknowledgments}

\bibliography{main_bib2}

\eject

\clearpage

\begin{figure}
\begin{center}
\vspace{-50mm}
\includegraphics[width=\textwidth]{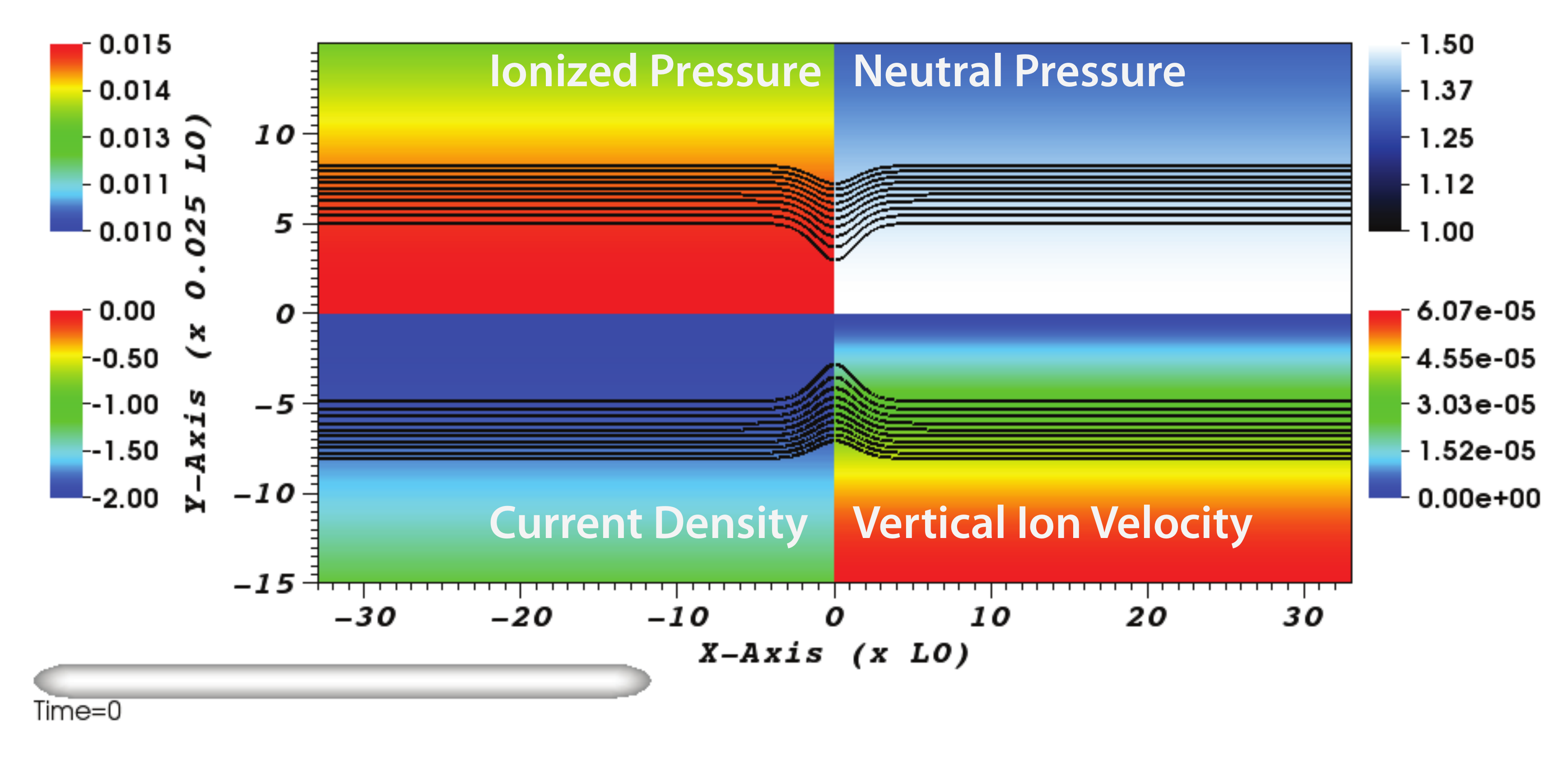}
\vspace{-10mm}
\caption{Initial conditions for a subset of the whole domain. Note that the $y$-coordinate is expanded  by a factor of 40. The top left quadrant shows ionized pressure ($(P_i+P_e)/P_{0}$), the top right quadrant shows neutral pressure ($P_{n}/P_{0}$), the bottom left quadrant shows current density ($j/(B_{0}/\mu_{0}L_{0})$), and the bottom right quadrant shows vertical ion velocity ($v_{i,y}/v_{0}$). The solid lines are 10 contour levels of the flux $A_{z}$ evenly distributed in the interval $[-0.04,-0.015]B_{0}L_{0}$.
\label{fig:IC1}}
\end{center}
\end{figure}


\begin{figure}
\begin{center}
\vspace{-0mm}
\includegraphics[width=\textwidth]{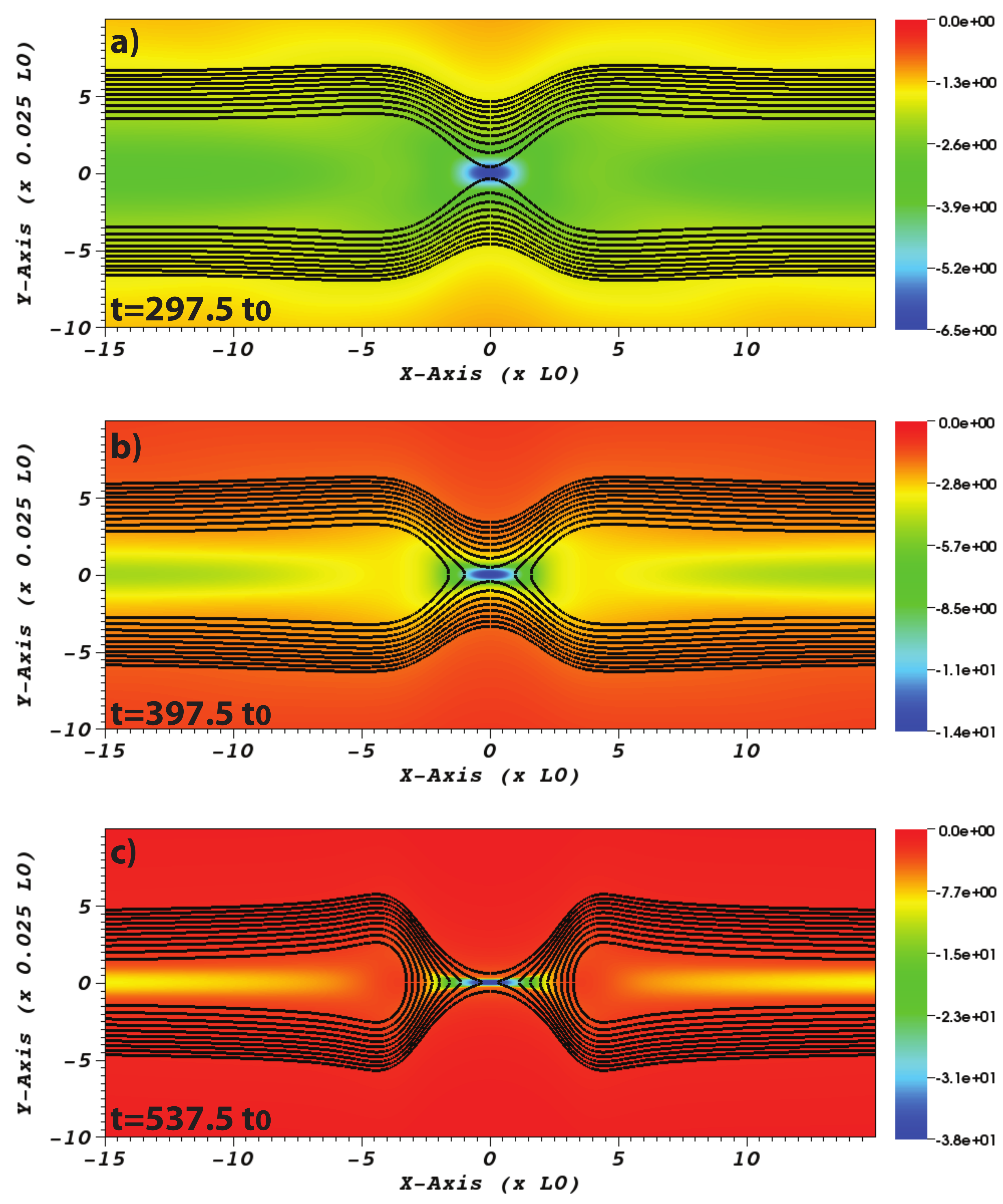}
\vspace{-10mm}
\caption{Formation of Sweet-Parker current sheet for the simulation where $\eta = 0.5\times 10^{-5}\eta_{0}$. Current density ($j/B_{0}/(\mu_{0}L_{0})$) is shown in the color contours, and 10 contour lines show $A_{z}$, evenly distributed in the interval $[-0.04, -0.015]B_{0}L_{0}$. \label{fig:reconnection}}
\end{center}
\end{figure}

\begin{figure}
\begin{center}
\vspace{-10mm}
\includegraphics[width=\textwidth]{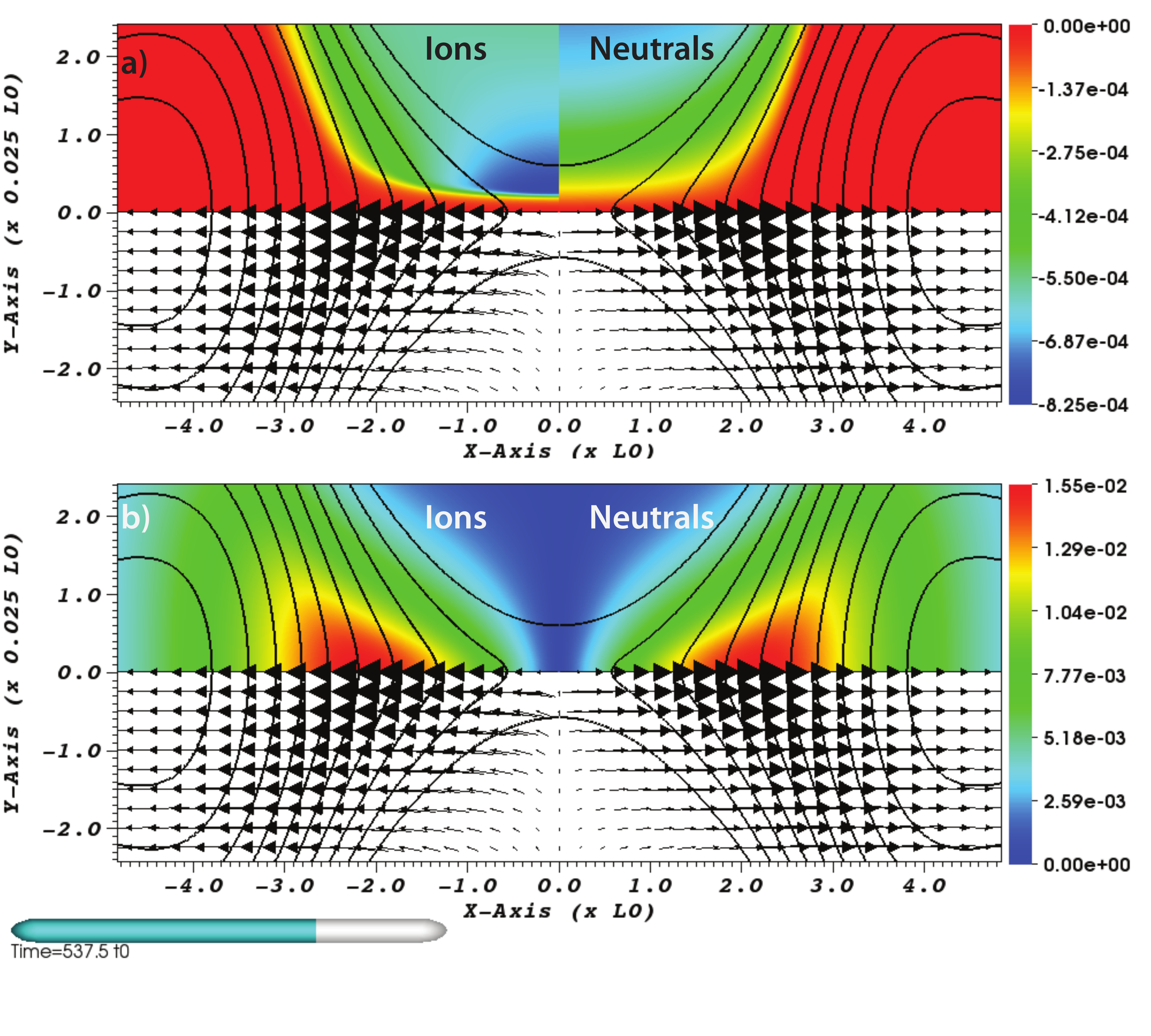}
\vspace{0mm}
\caption{Plots of ion and neutral flow, showing inflow decoupling and outflow coupling during reconnection. Panel a) shows color contours of the vertical velocity ($v_{i,y}/v_{0}$) for the ions (top left quadrant) and for the neutrals ($v_{n,y}/v_{0}$, top right quadrant). The contour lines show 10 values of $A_{z}$, regularly distributed in the interval $[-0.04,-0.01]B_{0}L_{0}$. The arrows on the bottom left quadrant represent the plasma flow, and those on the bottom right quadrant the neutral flow. Panel b) shows the same as panel a) but for the magnitude of horizontal velocity for ions ($|v_{i,x}/v_{0}|$) and neutrals ($|v_{n,x}/v_{0}|$). 
\label{fig:decoupling}}
\end{center}
\end{figure}

\begin{figure}
\begin{center}
\vspace{-0mm}
\includegraphics[width=\textwidth]{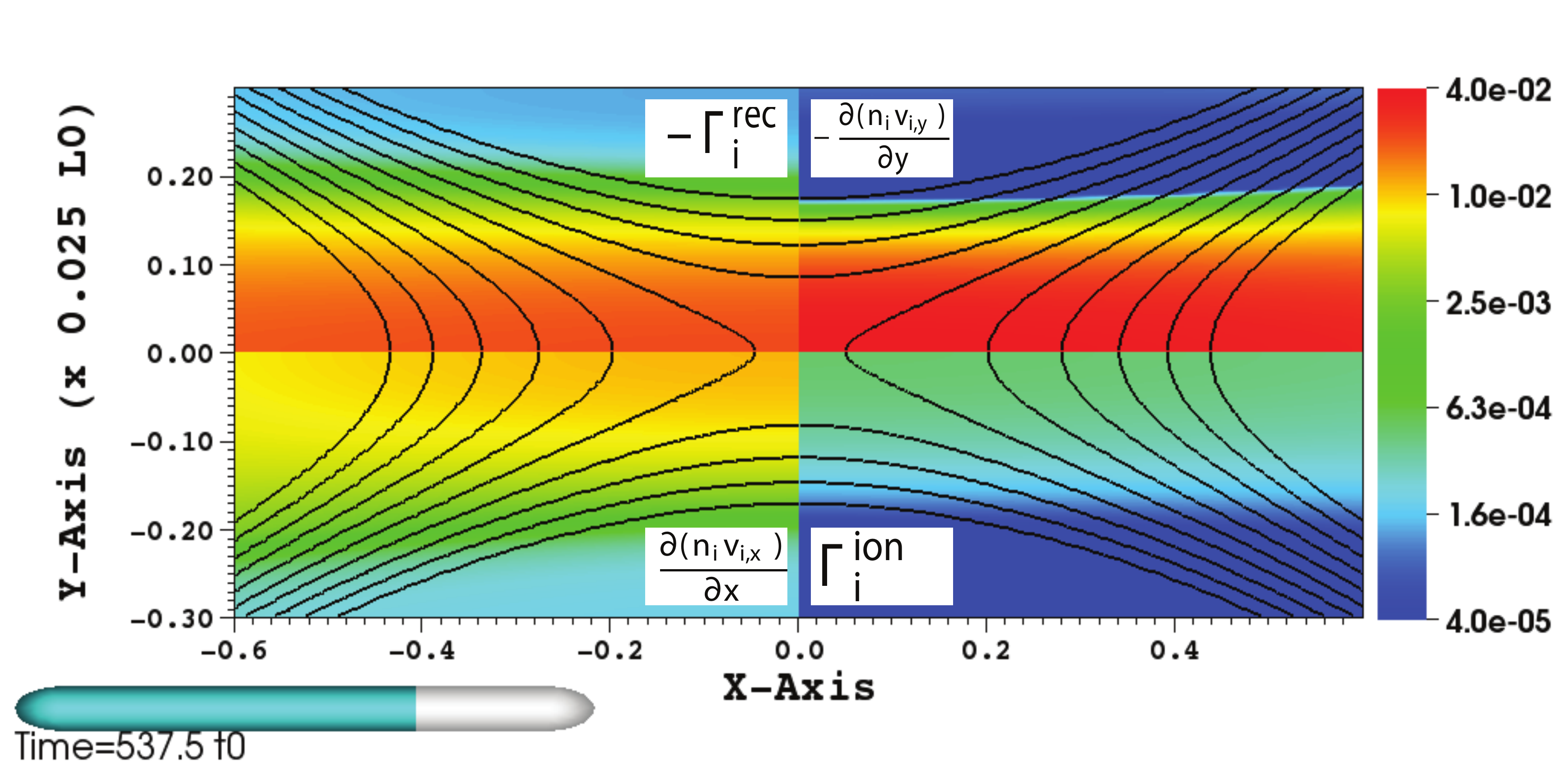}
\vspace{-0mm}
\caption{The steady state reconnection region showing contributing sources and sinks of ions in the current sheet (in units of $L_{0}^{-3}t_{0}^{-1}$): Top left quadrant shows rate of loss of ions due to recombination. Bottom left quadrant shows rate of loss of ions due to outflow $\frac{\partial n_{i}v_{i,x}}{\partial x}$. Top right shows rate of gain of ions due to inflow  $-\frac{\partial n_{i}v_{i,y}}{\partial y}$. Bottom right quadrant shows rate of gain of ions due to ionization. The solid lines are 10 contour lines of $A_{z}$, evenly distributed in the interval$ [-0.0378,-0.037]B_{0}L_{0}$. This shows that loss of ions due to recombination and outflow are comparable, and combine to balance the inflow of ions, with ionization playing an insignificant role.
\label{fig:steady_state}}
\end{center}
\end{figure}

\begin{figure}
\begin{center}
\vspace{-10mm}
\includegraphics[width=\textwidth]{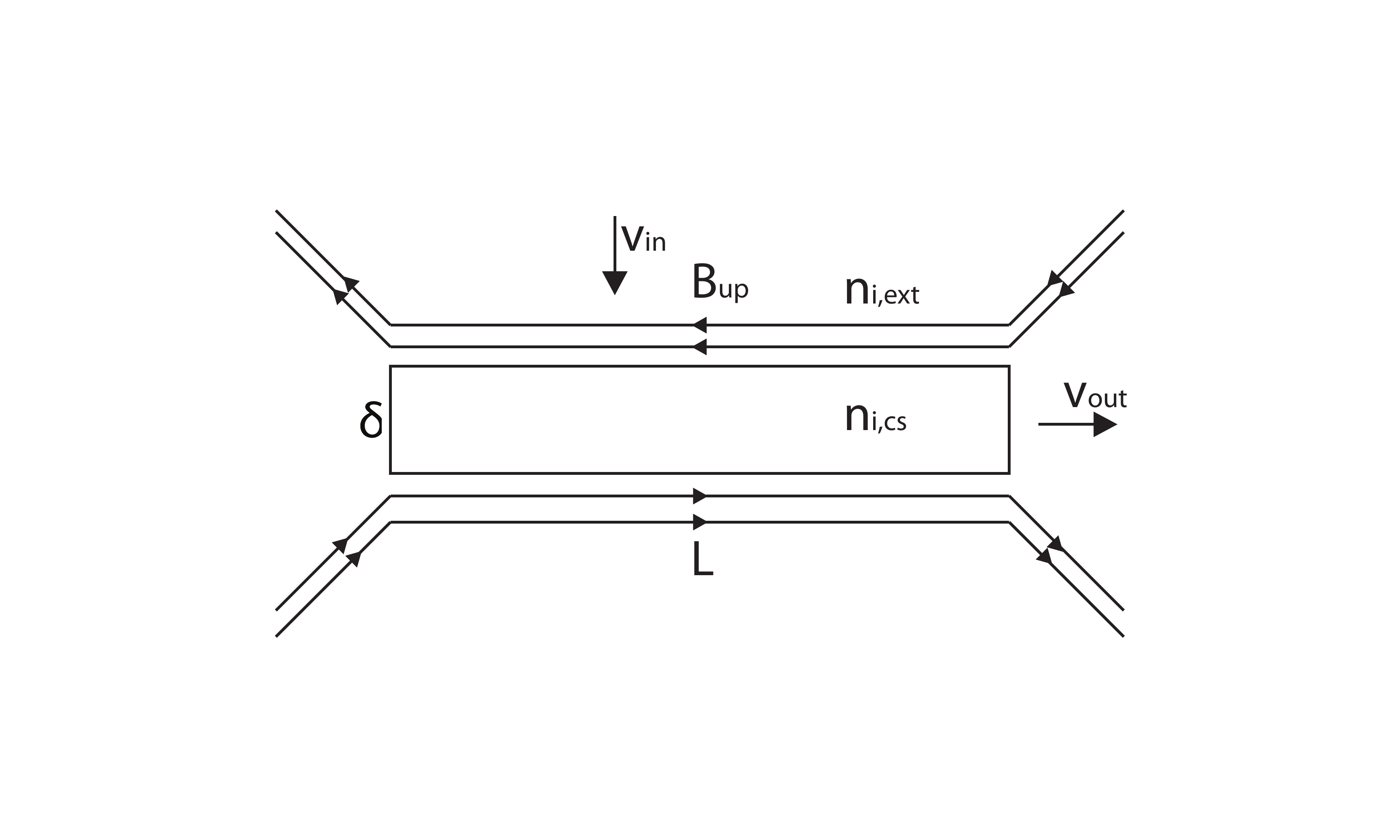}
\vspace{-20mm}
\caption{Cartoon of the Sweet-Parker reconnection scenario. The current sheet length is $L$ and width is $\delta$. The magnetic field just upstream from the current sheet is $B_{up}$. The ion inflow and outflow speeds are $v_{in}$ and $v_{out}$, respectively. The external and current sheet ion number densities are $n_{i,ext}$ and $n_{i,CS}$, respectively.
\label{fig:SP_PIP}}
\end{center}
\end{figure}

\begin{figure}
\begin{center}
\vspace{-10mm}
\includegraphics[width=\textwidth]{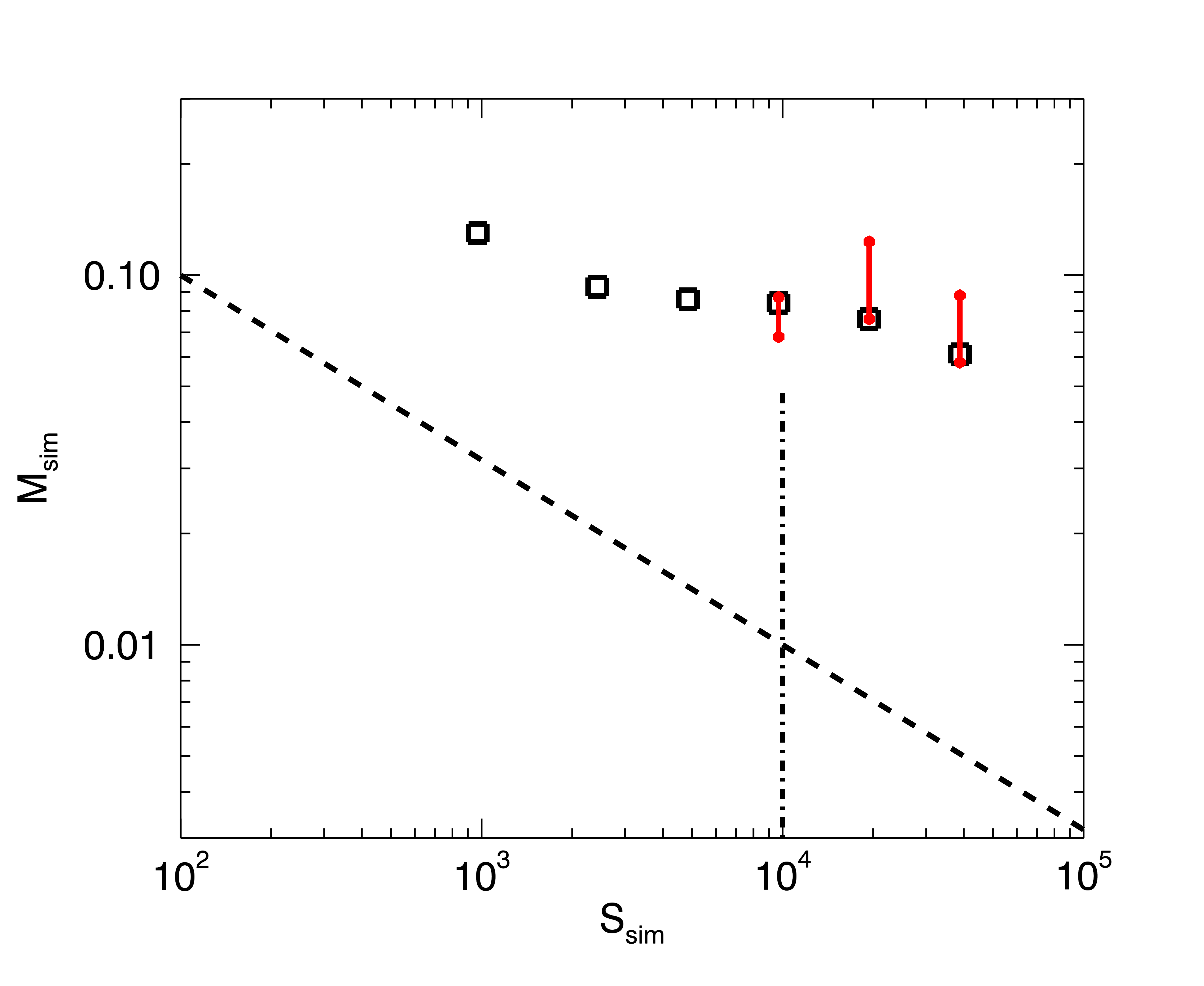}
\vspace{-10mm}
\caption{Normalized magnetic reconnection rate $M_{sim}$ for 6 simulations with different Lundquist number ($S_{sim}$). The squares show the reconnection rate taken at a time in each simulation when the length of the current sheet has reached $L_{sim}=2.75L_{0}$, and in all simulations is before the onset of the plasmoid instability. The red lines show the range in reconnection rate taken again at later times in the three plasmoid-unstable simulations, after the plasmoids are formed.
The dashed line is the Sweet-Parker scaling law $M \propto 1/\sqrt{S}$. The dot-dashed line shows the separation between plasmoid-stable and plasmoid-unstable regimes for these multi-fluid simulations, and is given by the value of $S_{sim}$ where the power law fit of the simulation data $\sigma_{sim}(S_{sim})$ meets the $\sigma_{sim}=1/200$ line in Figure \ref{fig:width}. 
\label{fig:rec_rate}}
\end{center}
\end{figure}

\begin{figure}
\begin{center}
\vspace{-10mm}
\includegraphics[width=\textwidth]{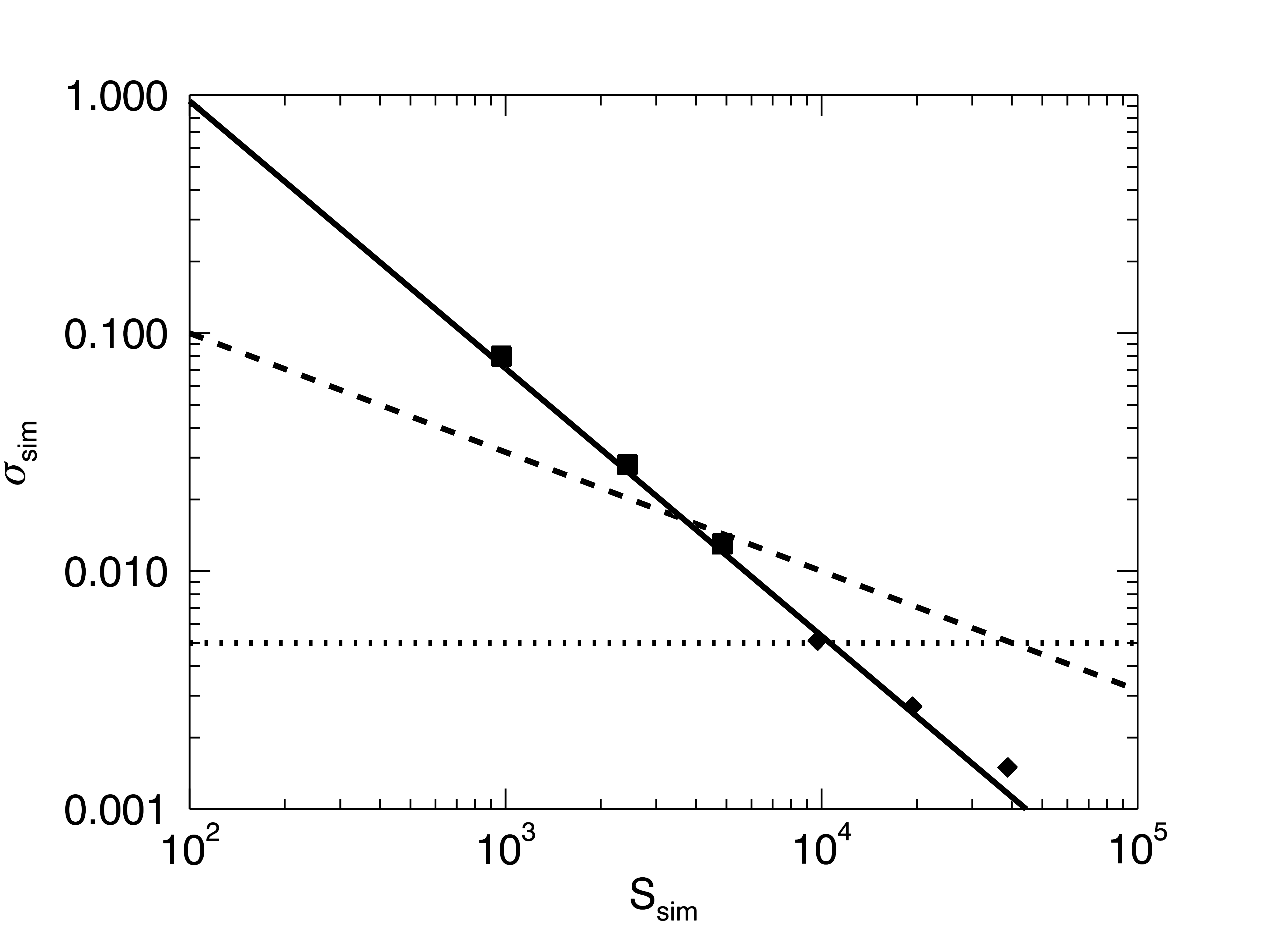}
\vspace{-10mm}
\caption{Scaling of simulation current sheet aspect ratio $\sigma_{sim}$ with Lundquist number ($S_{sim}$). The squares are simulations where no secondary (plasmoid) instability is seen, and the diamonds are simulations where plasmoids are observed. The dotted line is the theoretical aspect ratio at which the plasmoid instability sets in. The dashed line is the Sweet-Parker scaling law ($\propto \sqrt{1/S}$). The solid line shows a line of best fit of the data to a power law. The exponent in the power law is $-1.1 \pm 0.17$. This solid line intersects the $\sigma_{sim}=1/200$ line at approximately $S_{sim}=10^{4}$.
\label{fig:width}}
\end{center}
\end{figure}

\begin{figure}
\begin{center}
\vspace{-0mm}
\includegraphics[width=\textwidth]{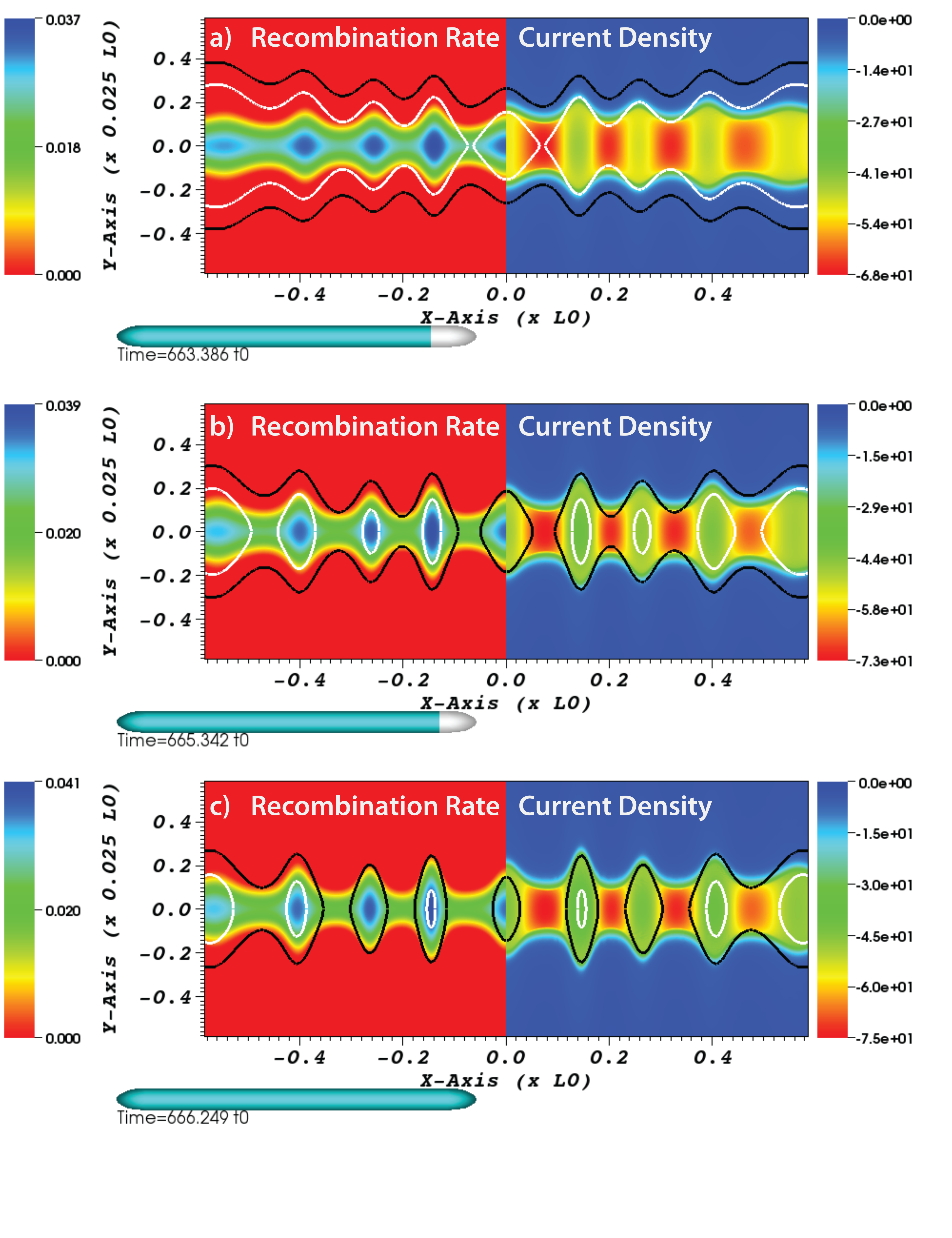}
\vspace{-20mm}
\caption{Plasmoid formation and evolution: Recombination rate $\Gamma_{i}^{rec}L_{0}^{3}t_{0}$ (left) and current density $j\mu_{0}L_{0}/B_{0}$ (right), and  two contour levels of $A_{z}$ of $-0.0681B_{0}L_{0}$ (white) and $-0.0687B_{0}L_{0}$ (black), at three different times in the simulation where $\eta=0.5\times10^{-5}\eta_{0}$.
\label{fig:plasmoid}}
\end{center}
\end{figure}

\end{document}